\documentclass[11pt,twocolumn]{article}
\usepackage{graphicx}
\usepackage[colorlinks=true]{hyperref}
\usepackage[dvipsnames]{xcolor}
\usepackage{wrapfig}
\usepackage{setspace}
\usepackage{natbib}
\usepackage{textcomp}
\usepackage{datetime}
\usepackage{authblk}
\usepackage[]{fancyhdr}

\hypersetup{
    colorlinks=true,
    linkcolor=RoyalBlue,
    filecolor=magenta,      
    urlcolor=cyan,
    pdftitle={Overleaf Example},
    pdfpagemode=FullScreen,
    citecolor=Magenta,
    }

\shortdate
\settimeformat{hhmmsstime} 

\newcommand{\bof}[1]{}
\newcounter{NTRB}
\newcounter{NTRC}
\newcounter{NTRA}
\definecolor{jxlCa}{HTML}{5882FF}

\let\OLDthebibliography\thebibliography
\renewcommand\thebibliography[1]{
  \OLDthebibliography{#1}
  \setlength{\parskip}{0pt}
  \setlength{\itemsep}{5pt plus 0.3ex}
}

\newcommand{\be}{\begin{equation}}
\newcommand{\ee}{\end{equation}}

\def\m@th{\mathsurround=0pt}
\newcommand\EQM[1]{\vcenter{\normalbaselines\m@th
    \ialign{${\displaystyle ##}$\hfil&&\ ${\displaystyle ##}$\hfil\crcr
    \mathstrut\crcr\noalign{\kern-\baselineskip}
    \noalign{\smallskip}
    #1\crcr\mathstrut\crcr\noalign{\kern-\baselineskip}}}}

\newcommand\Frac[2]{{{\displaystyle\strut#1}\over{\displaystyle\strut#2}}}

\def\temi{\Frac{3}{2}}
\def\moo{M} 
\def\ed{{E\!_d}}
\def\mo{m_\odot}
\def\o{\odot}

\newcommand\figa{
\begin{figure}[t!]
\centering
  \includegraphics[width=0.9\linewidth]{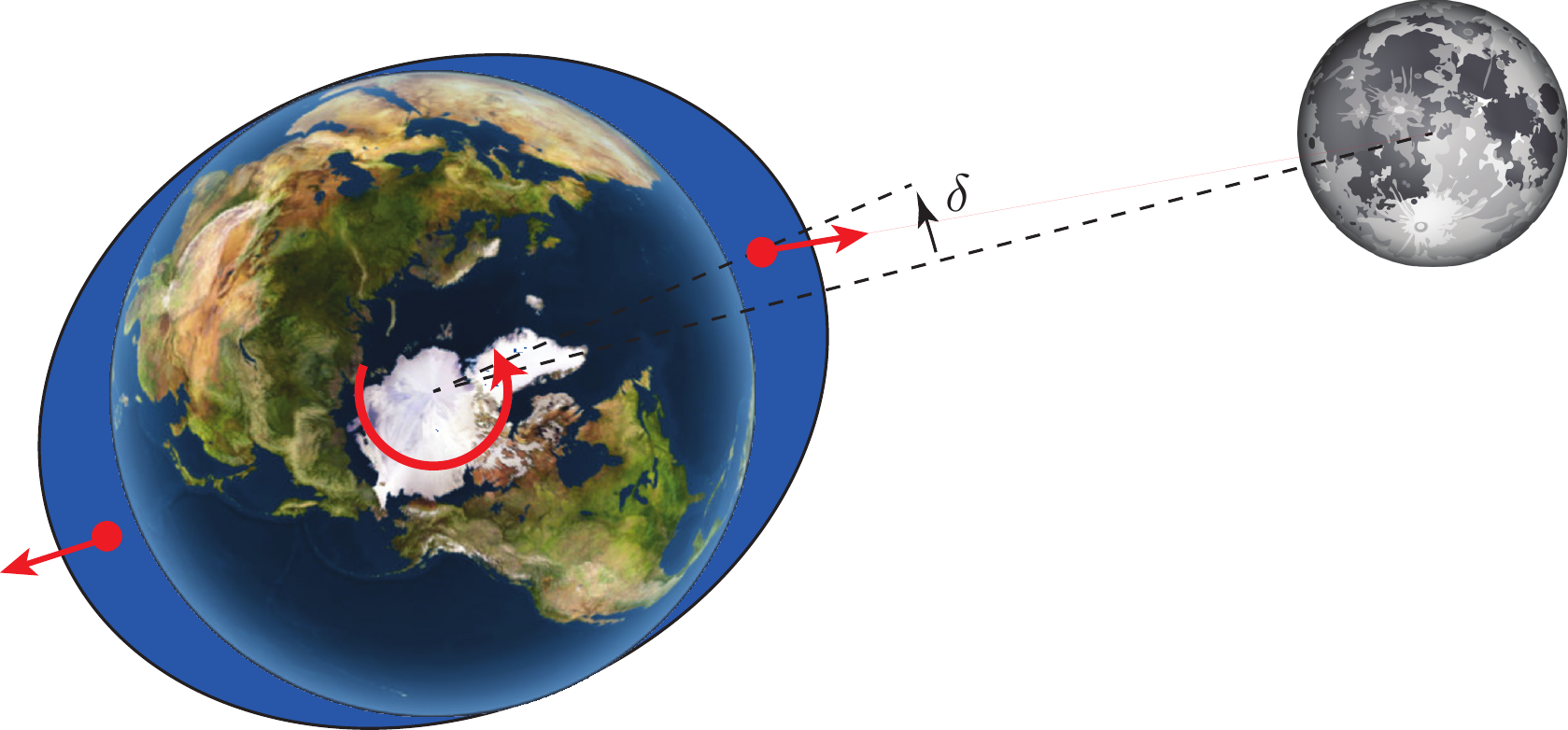}
  \caption{\footnotesize In Darwin model, the Moon produces a tidal bulge on the Earth, but as the Earth is not totally elastic, this bulge is driven by the fast rotation of the Earth slightly off from the Moon direction. From this results a braking torque that slows  down the spin of the Earth. By conservation of the angular momentum in the Earth-Moon system, the Moon goes away. }
  \label{fig:darwin}
\end{figure}
}

\newcommand\figb{
\begin{figure}[h!]
  \includegraphics[width=\linewidth]{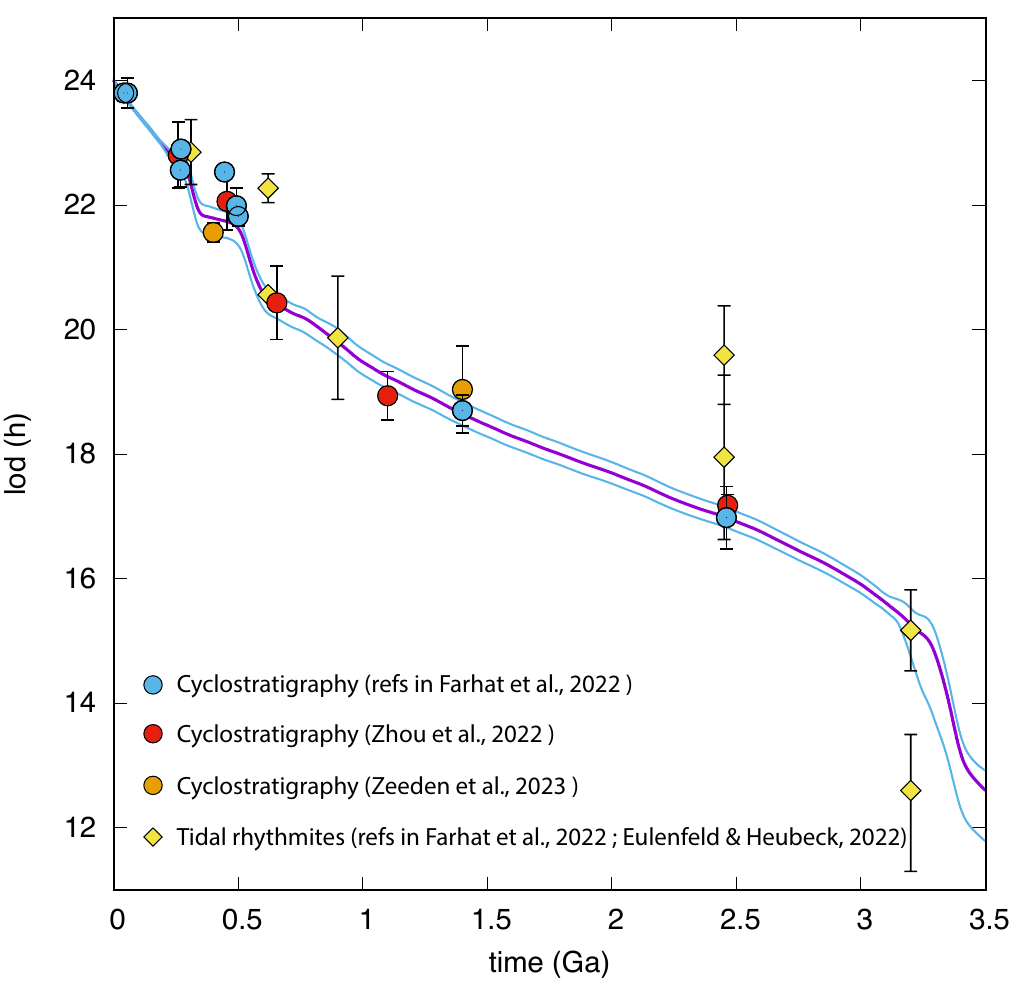}
  \caption{\footnotesize 
  Evolution of the length of day (LOD) in the past following the model of  \citep{farhat2022resonant}. The nominal LOD values are in purple, with the uncertainty provided by the blue lines (adapted from fig.5 of \citep{farhat2022resonant}).
  The circles represent the available cyclostratigraphic data with their uncertainty, when available, from various sources:  references within \citep{farhat2022resonant} (in blue), 
  \citep{ZhouWu2022a} (in red), and \citep{ZeedenLaskar2023a} (in orange).
  Tidal rhythmites values are represented by yellow squares, with references from 
   \citep{farhat2022resonant} with the addition of a data point at 3.2 Ga from the Moodies group \citep{eulenfeld_constraints_2023}.
}
  \label{fig:f22}
\end{figure}
}

\newcommand\figc{
\begin{figure}[t!]
\centering
  \includegraphics[width=0.9\linewidth]{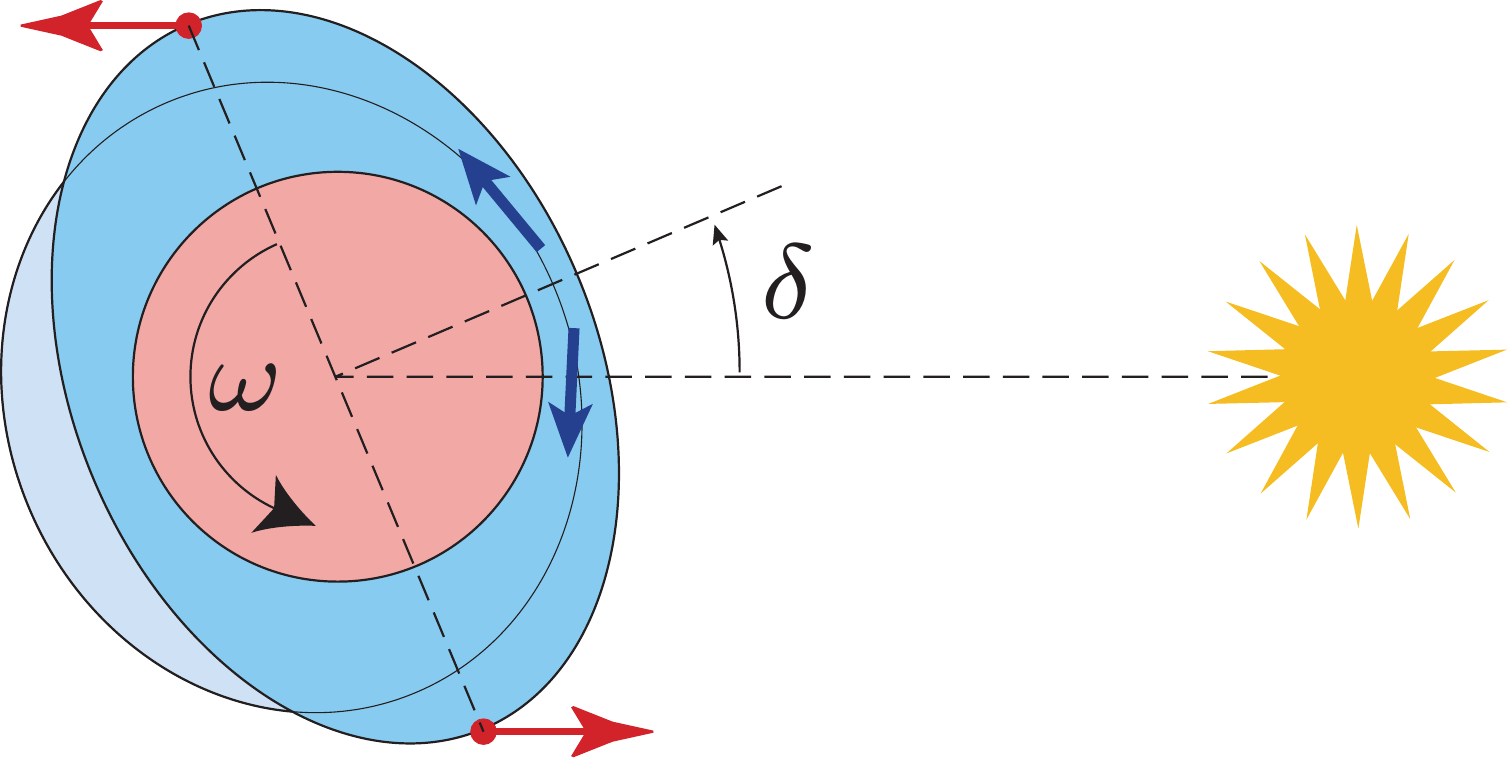}
  \caption{\footnotesize Thermal atmospheric tides. Due to the heating of the Sun at the subsolar point, a redistribution of the atmosphere creates two main components:  a diurnal component (in light blue), and a semi diurnal components (in blue) that are offset from their equilibrium position because of the Earth's fast rotation. They create an accelerating torque on the Earth's spin motion.}
  \label{fig:TAT}
\end{figure}
}

\newcommand\figd{
\begin{figure}[h!]
  \includegraphics[width=\linewidth]{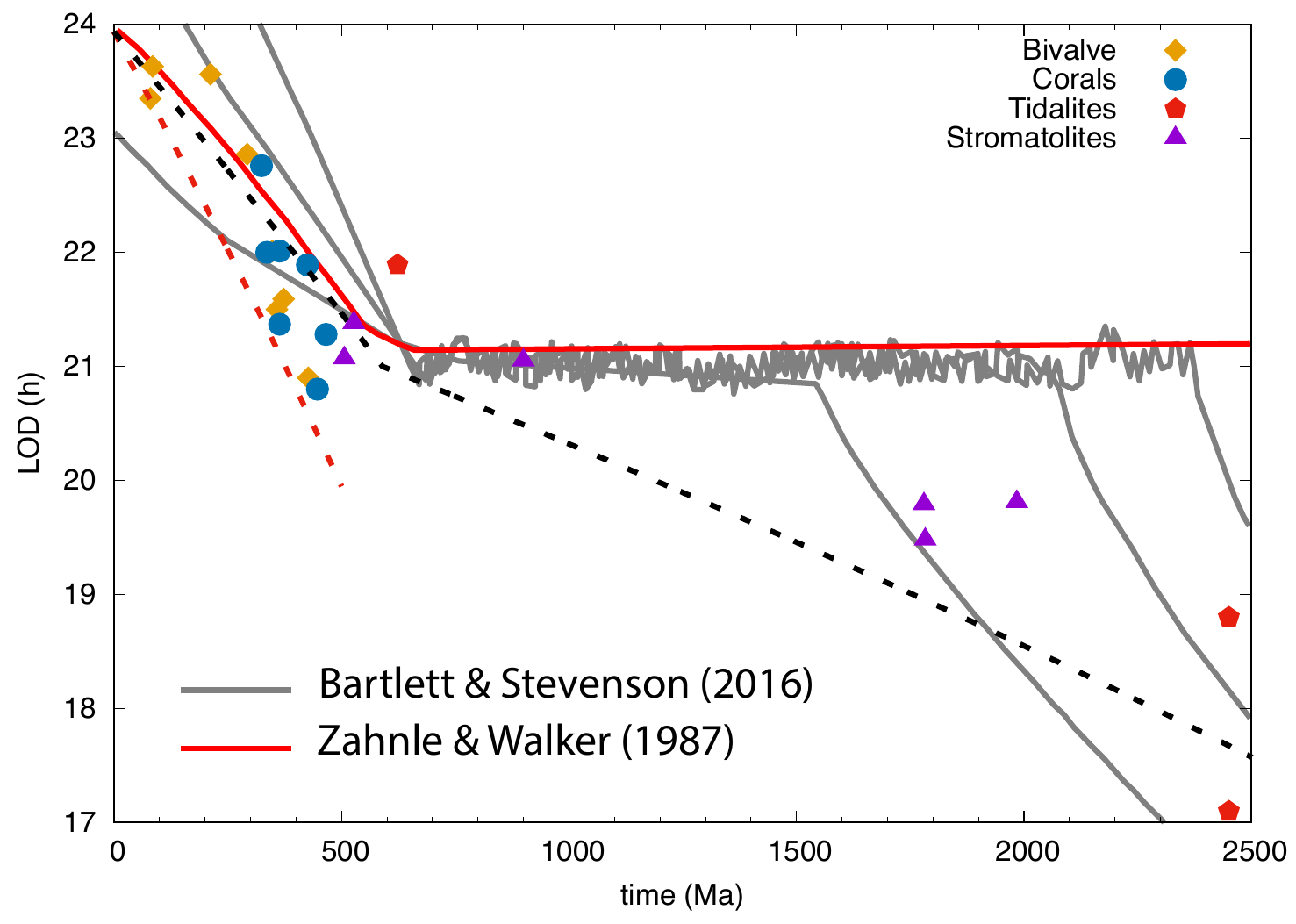}
  \caption{\footnotesize LOD locking due to thermal atmospheric tides 
  resonance for the scenarios of \citep{zahnle1987constant} (in red)
and \citep{bartlett2016analysis} (in grey). The geological indicators that are plotted are limited to pre-2016 published results, adapted from \citep{Laskar2020a} and \citep{williams2000geological} (see references therein).    The  dotted red line is the LOD provided by equation 41 of
\citep{laskar2004long} with a simple Darwin tidal model. The dotted black line is an empirical fit using a simplified tidal model adjusted to the geological data \citep{walker1986lunar}.
The stromatolite data points at 1.88 and 2.0 Ga are from \citep{pannella1972paleontological,pannella1972precambrianB}.
}
  \label{fig:LOD}
\end{figure}
}

\newcommand\fige{
\begin{figure*}[h!]
  \includegraphics[width=\linewidth]{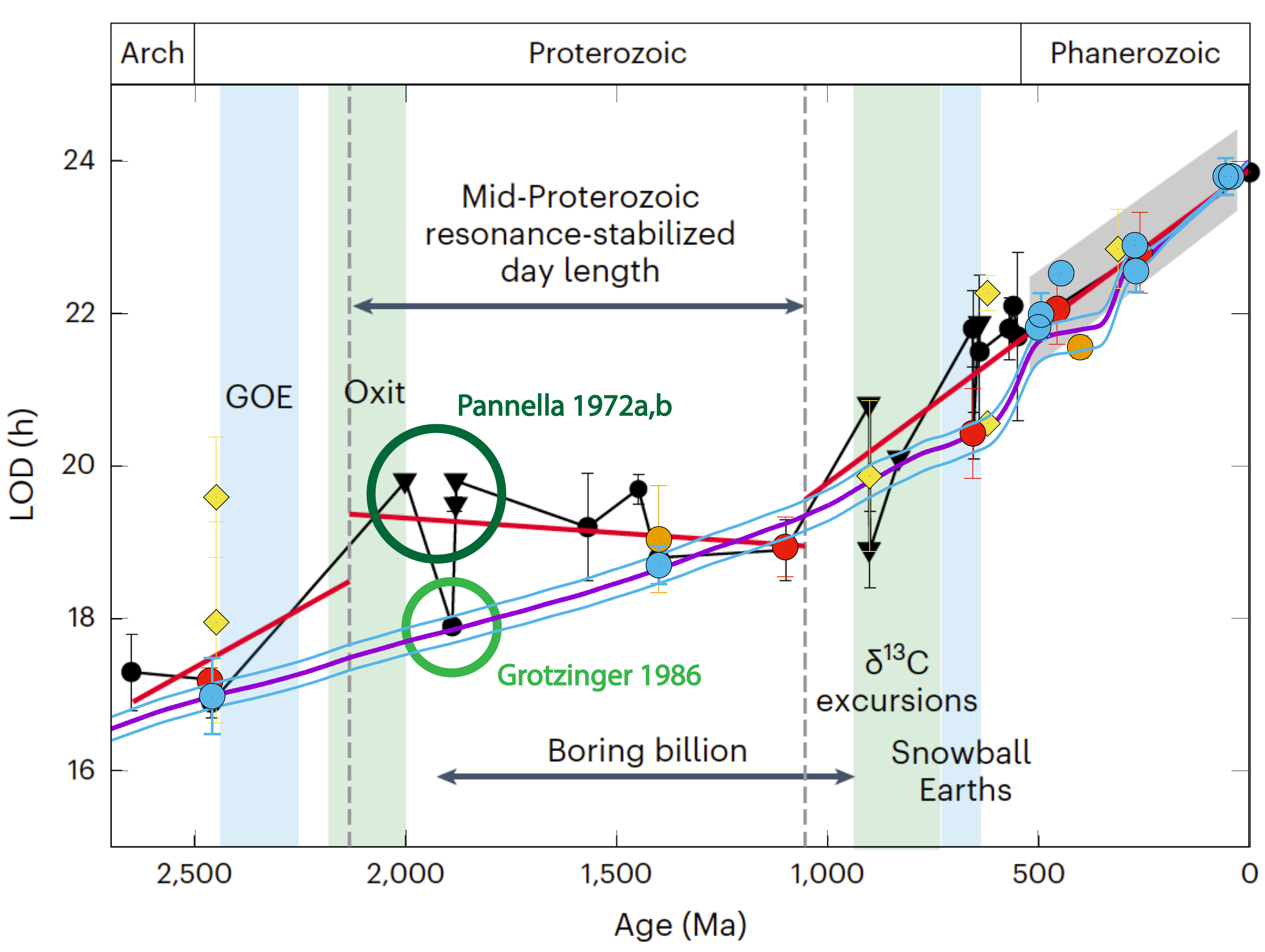}
  \caption{\footnotesize 
  Comparison of the data and fit of \citep{Mitchell2023a}  (black and red lines) with the model of \citep{farhat2022resonant}  (purple curve with uncertainty in blue). As 
 in Fig.\ref{fig:f22}, tidal rhythmites are yellow squares, and cyclostratigraphic data are color circles (light blue with references in  \citep{farhat2022resonant}, red from  \citep{ZhouWu2022a} and orange from \citep{ZeedenLaskar2023a}). The stromatolite data from \citep{pannella1972paleontological,pannella1972precambrianB} are highlighted with a dark green circle while the cyclostratigraphic data from \citep{grotzinger_cyclicity_1986} is circled in light green. Adapted from Fig.2 of \citep{Mitchell2023a}. 
}
  \label{fig:mitchell}
\end{figure*}
}

\newcommand\figf{
\begin{figure*}[t!]
  \includegraphics[width=\linewidth]{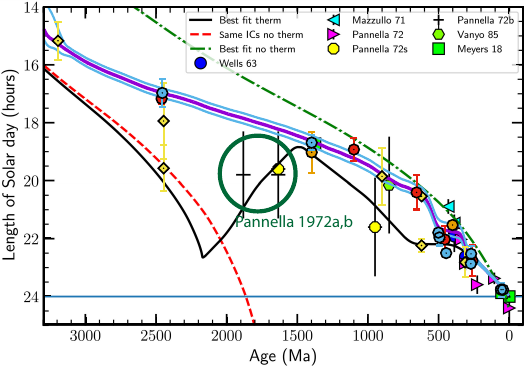}
  \caption{\footnotesize 
  Comparison of the data and fit of \citep{wu2023day}  (in black) with the model of \citep{farhat2022resonant}  (purple curve with uncertainty in blue). As 
 in Fig.\ref{fig:f22}, tidal rhythmites are yellow squares, and cyclostratigraphic data are color circles (light blue with references in  \citep{farhat2022resonant}, red from  \citep{ZhouWu2022a} and orange from \citep{ZeedenLaskar2023a}). The stromatolite data from \citep{pannella1972paleontological,pannella1972precambrianB} are highlighted with a dark green circle. Note that these data points seem to be misplaced in the \citep{wu2023day} figure reproduced here (see note 1).
 Adapted from Fig.3 of  \citep{wu2023day}.
}
  \label{fig:wu}
\end{figure*}
}

\newcommand\figg{
\begin{figure}[h!]
  \includegraphics[width=\linewidth]{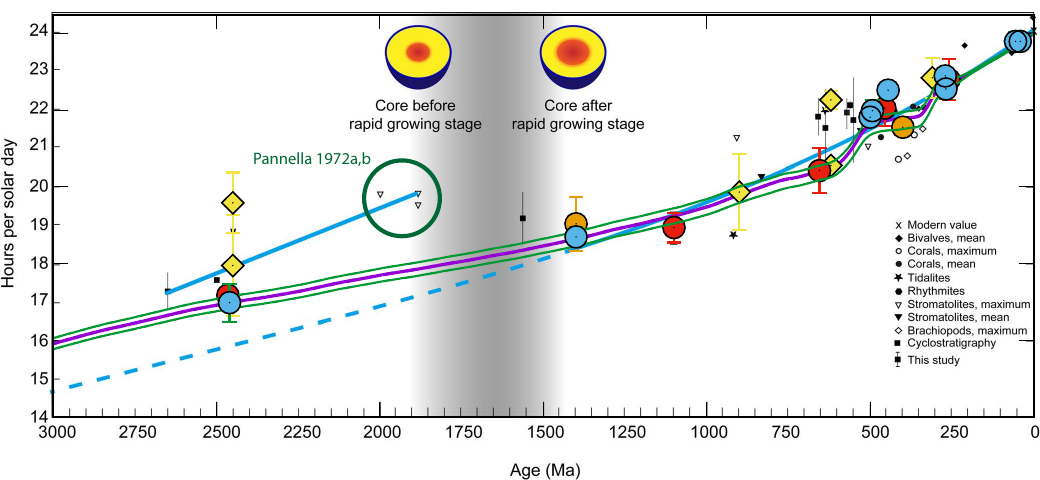}
  \caption{\footnotesize 
  Comparison of the data and fit of \citep{baolength2023} 
  (in solid blue) with the model of \citep{farhat2022resonant}  (purple curve with uncertainty in green). As 
 in Fig.\ref{fig:f22}, tidal rhythmites are yellow squares, and cyclostratigraphic data are color circles (light blue with references in  \citep{farhat2022resonant}, red from  \citep{ZhouWu2022a} and orange from \citep{ZeedenLaskar2023a}). The stromatolite data from \citep{pannella1972paleontological,pannella1972precambrianB} are highlighted with a dark green circle.
 Adapted from Fig.11 of \citep{baolength2023}.
  \label{fig:bao}
  }
\end{figure}
}

\newcommand\figh{
\begin{figure}[h!]
  \includegraphics[width=\linewidth]{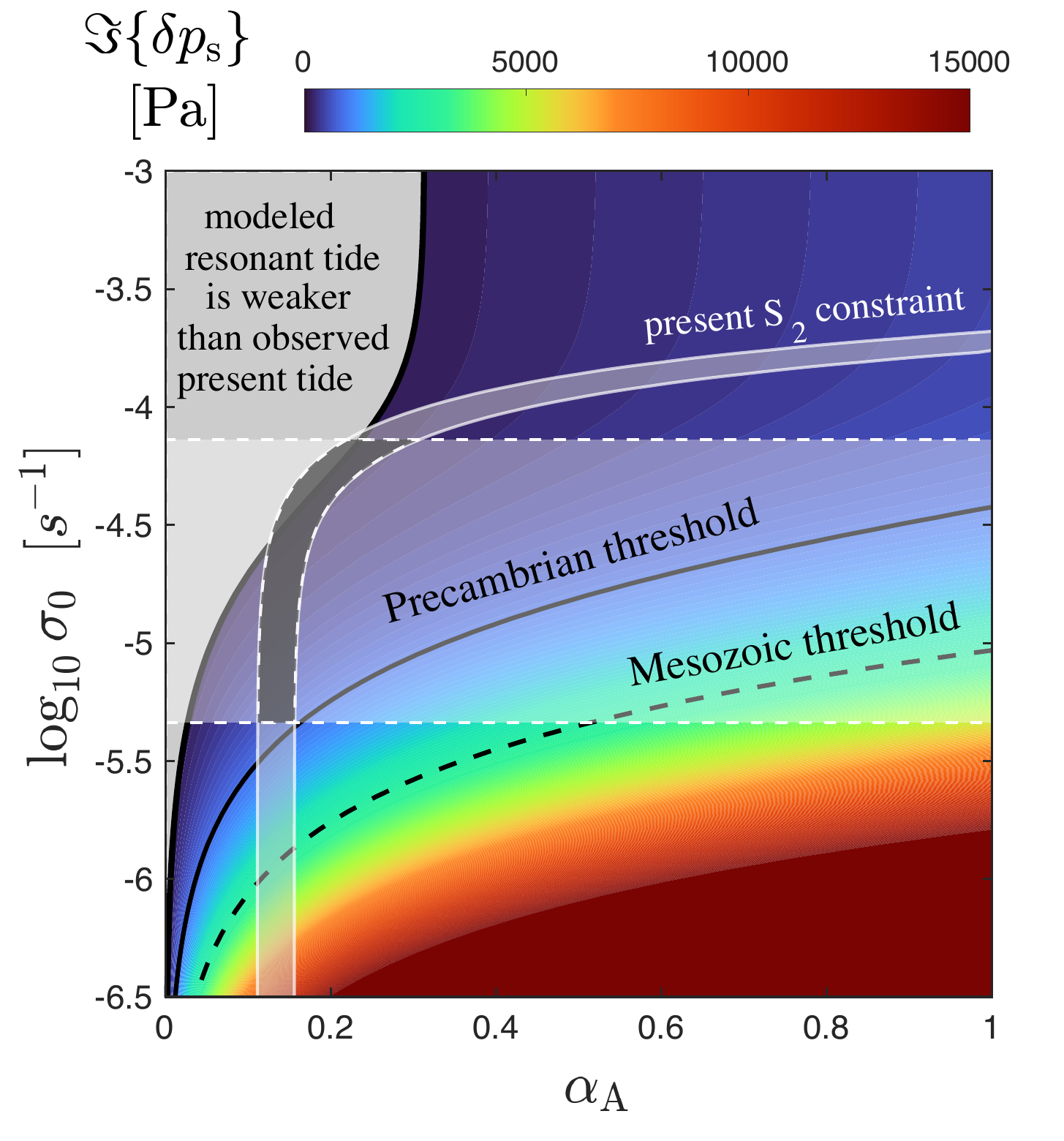}
  \caption{\footnotesize 
 Amplitude of the Lamb resonance with respect to the two parameters $\sigma_0$ and $\alpha_{\mathrm{A}}$ of the \citep{FarhatAuclair-Desrotour2023a} model.  In order for the thermotidal response  to cancel the gravitational counterpart in the Precambrian the parameters $(\sigma_0, \alpha_{\mathrm{A}})$ need to be below  the solid black line. 
 The dashed line defines the same threshold needed for the Mesozoic, at 250 Ma. The horizontal shaded area corresponds to typical values of the radiative cooling rate  $\sigma_0$. The other shaded area defines the region of parameter space corresponding to the presently observed semi-diurnal tidal bulge.
 Adapted from \citep{FarhatAuclair-Desrotour2023a}.}
  \label{fig:fat}
\end{figure}
}

\newcommand\figi{
\begin{figure}[h!]
  \includegraphics[width=\linewidth]{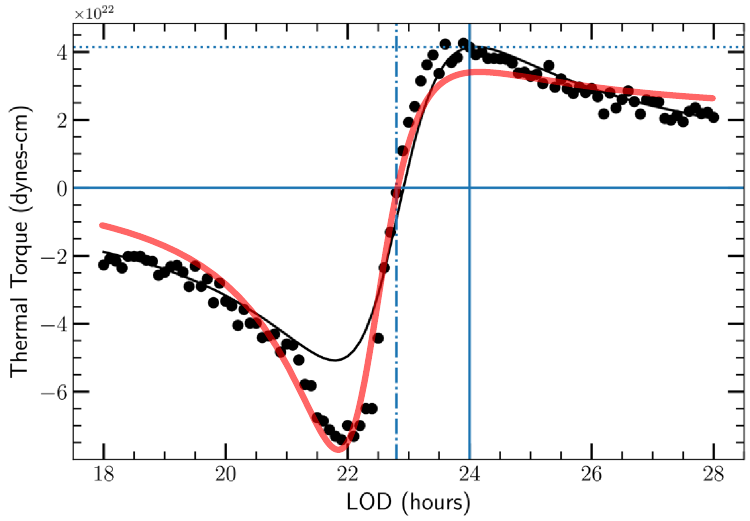}
  \caption{\footnotesize  The spectrum of the  thermotidal torque as a function of the length of day adapted from Fig.S4 of \citep{wu2023day}. The black dots are simulated using the PlaSim GCM, while the solid black curve shows the prediction of the analytical model of \cite{wu2023day}. The red curve shows a fit to the GCM results using the analytical model of \cite{FarhatAuclair-Desrotour2023a}, where the physical effects of the delayed thermal response of the ground were taken into account. These effects proved to induce the notable amplitude asymmetry between the peaks around the Lamb resonance. Namely, the accelerating peak of the tidal torque is reduced in amplitude relative to the decelerating peak.
 }
  \label{fig:Wu_Spectrum}
\end{figure}
}

\title{\bfseries Did atmospheric thermal  tides cause  a 
 daylength locking\\ in the Precambrian? \\ A review on recent results.}
\author[1*]{Jacques Laskar}
\author[1]{Mohammad Farhat}
\author[2]{Margriet L. Lantink}
\author[1]{Pierre Auclair-Desrotour}
\author[1]{Gwena\"el Bou\'e}
\author[1]{Matthias Sinnesael}

\affil[1]{IMCCE, CNRS, Observatoire de Paris, PSL University, Sorbonne Université, \\
77 Av. Denfert-Rochereau 75014, Paris, France}
\affil[2]{Department of Geoscience, University of Wisconsin–Madison, Madison, WI 53706, Unites States. \vspace{1cm}}

\affil[*]{Corresponding author: Jacques Laskar (jacques.laskar@obspm.fr)}

\begin{document}
\bof{
\phantom{0}
\vspace{-1 cm}
\begin{center}
{\LARGE
{\Huge Jacques Laskar }\\[1cm] 
Notice \\[0.5 cm]  
\Large 26 octobre 2018 \\[1cm] 
}
\end{center}

\clearpage
}
\thispagestyle{empty}
\medskip

\bof{
\twocolumn[%
\section*{Did atmospheric thermal  tides cause a 
 daylength locking in the Precambrian? \\ A review on recent results.}
\noindent
Authors: J. Laskar, M. Farhat, M.  Lantink, P.auclair-Desrotour, G. Boué, M. Sinnesael
\medskip
\noindent 
\begin{abstract}


\noindent 
{\bf Key words:} Milankovitch cycles -- Thermal atmospheric tides -- Earth-Moon distance -- Tidal friction -- Precambrian Earth-- LOD resonant locking.
\end{abstract}%
]
}

\twocolumn[
\maketitle

\medskip
\hrule

\begin{abstract}
  After the initial suggestion by Zahnle and Walker (1987) that the torque accelerating 
  the spin rate of the Earth and 
  produced by the heating of the atmosphere by the Sun could counteract the braking lunir-solar gravitational torque in the Precambrian, several authors have recently revisited  this 
  hypothesis. In these studies, it is argued that the geological evidences of the past spin state of the Earth 
  play in favor of this atmospheric tidal locking of the  length of the day (LOD). 
  In the present review of the recent literature, we show that the drawn conclusions depend crucially on the consideration of the stromatolite  geological LOD estimates  obtained by Pannella at 1.88 and 2.0 Ga,  which are subject to large uncertainties. When only the most robust cyclostatigraphic estimates of the LOD are retained, the LOD locking 
  hypothesis is not supported. Moreover, the consideration of the published General Circulation Model numerical simulations and of new analytical models for the thermal atmospheric tides suggest that the atmospheric tidal resonance, 
 which is the crucial ingredient for the LOD locking in the Precambrian, was never of sufficiently 
  large amplitude to allow for this tidal LOD lock. 
\end{abstract}

\medskip
\noindent 
{\bf Key words:} Milankovitch cycles -- Thermal atmospheric tides -- Earth-Moon distance -- Tidal friction -- Precambrian Earth-- LOD resonant locking.

\medskip
\hrule
\medskip
\phantom{0}
\vspace{5 cm}
]


\tableofcontents
\section{Introduction}
\label{sec:intro}
Since the work of Georges Darwin,  it is known that the body tides exerted by the Sun and Moon on Earth slow down the spin of the Earth and make the Moon recede away  (Fig.1) \citep{Darwin1879,macdonald_tidal_1964,goldreich1966history,kaula_tidal_1964,mignard_evolution_1979,mignard1980evolution,hut1981tidal,touma_evolution_1994,NeroLask1997a}. 
\figa
More elaborate tidal models take into account the oceanic tides which also slow down the rotation of the Earth and let the Moon go away \citep{webb1982tides,green2017explicitly,tyler2021tidal,daher2021long}, but none of these tidal models could fit both the present tidal recession of the Moon of $3.83 \pm 0.008$ cm/yr \citep{williams2016secular} and the age of the Moon of $4.425 \pm 0.025$ Ga \citep{maurice2020long}. Elaborated along the lines of \citep{webb1982tides}, the recent semi-analytical model  of \citep{farhat2022resonant} provides a coherent scenario for the Earth-Moon tidal evolution, with an excellent fit to the present recession rate and the age of the Moon. 
It starts with a global ocean in the ancient eons, and then switches to a hemispheric ocean model similar to \citep{webb1982tides}, but which follows the continental evolution in the most recent times.
Although no geological data was used for the elaboration of the model, it is independently in good agreement with   geological estimates of the past Earth-Moon distance, and in particular of the geological constraints obtained by cyclostratigraphic methods (Fig.2). Nevertheless, following the original suggestion of \citep{zahnle1987constant}, two recent papers propose that the spin of the Earth was trapped in a resonance between the thermal atmospheric torque and the solid and oceanic torque during the Precambrian Eon \citep{Mitchell2023a,wu2023day}. Both studies advocate that this locking of Precambrian daylength is supported by geological observations. However, such a scenario is incompatible with the Earth-Moon evolution presented in \citep{farhat2022resonant}. In this review, which is aimed to the stratigraphic community, we will explicit  the differences in the modelling  approaches and discuss the use of geological data that can be compared with the models.  
\figb

\subsection{Atmospheric thermal tides}
\label{sec:TAT}

 Atmospheric thermal tides have been recognized since the XVIII th century \citep[see][]{Wilkes49}. Due to the heating of the atmosphere by the Sun, the atmosphere locally expands and the pressure decreases at the subsolar point, which induces a redistribution of the mass of the atmosphere, with two main components: a diurnal component, which at equilibrium is opposite to the subsolar point, and a semi-diurnal component, orthogonal to the Sun direction (Fig.3). As for the solid tides, as the Earth spin rotation is faster than its orbital motion around the Sun, the Earth rotation drags these atmospheric bulges with a positive offset from their equilibrium position. At present, the gravitational attraction of the Sun on these bulges induces an accelerating torque, opposite to the solid and ocean tides \citep{chapman1969atmospheric,
goldreich1966q,gold1969atmospheric,ingersoll1978venus,dobrovolskis1980atmospheric,correia2001four,correia2003long2,correia2003long,leconte2015asynchronous,auclair2017atmospheric,auclair2019generic}.

\figc

\subsection{A possible lock of the length of the day}
\label{sec:LOD}
\figd
At present, the thermal atmospheric tidal torque  is a small part ($\sim 6.4$ \%) of the solid and oceanic tidal friction  \citep{volland_atmospheric_1990,FarhatAuclair-Desrotour2023a}, but there are two elements that can change this ratio, in favor of the atmospheric thermal tides. First, the atmospheric torque amplitude is dependent on the spin rate of the Earth, and as for the oceanic tides, there exists a known resonance of the planetary Lamb wave \citep{bretherton_lamb_1969}, that we name here the Lamb resonance \citep{FarhatAuclair-Desrotour2023a}, occurring for a faster spin value of the Earth, where the atmospheric torque is largely increased \citep{lindzen1972lamb,zahnle1987constant,bartlett2016analysis}. In addition, the oceanic tidal friction is  at present close to a resonance, but its value was smaller in the past, for a faster rotation spin value \citep[e.g.][]{farhat2022resonant}. These elements led \citet{zahnle1987constant} to  propose that at some time  in the Precambrian, the accelerating thermotidal torque couter-acted the braking luni-solar gravitational tidal torque, which led to a lock of the length of the day (LOD)  at about 21h for an extended period of more than 1 Ga.  This assumption was probably motivated by the difficulty to fit the existing geological indicators of the past Earth-Moon distance and LOD with more simple models \citep[e.g.][]{williams2000geological,Hinn2018b,Laskar2020a}. 
This hypothesis was recently revisited by \citet{bartlett2016analysis}.

Both in \citep{zahnle1987constant} and \citep{bartlett2016analysis}, 
a large part of the observational constraints was provided from the Precambrian estimates of the LOD resulting from stromatolites deposit analysis 
\citep{pannella1972paleontological,pannella1972precambrianB} (Fig.\ref{fig:LOD}). 

\section{Geological archives for Precambrian LOD estimates}
In  his review on geological archives for LOD estimates, \citet{williams2000geological}
mentioned several varieties of bio-archives (bivalves, corals, brachiopods) which we will not consider here, as they  were not  present in the Precambrian. 
For these, one can also consult \citep{rosenberg_growth_1975} or \citep{lambeck2005earth}.
We will concentrate here on the stromatolites (\ref{sec:stro}), the tidal rhythmites (\ref{sec:rhy}), and the promising cyclostratigraphic records (\ref{sec:cycl}).

\subsection{Stromatolites}
\label{sec:stro}
The model of \citep{bartlett2016analysis} relies heavily on the adjustment to stromatolite data \citep{pannella1972paleontological,pannella1972precambrianB}, although the authors themselves warn the reader, and we can  quote from \citep{bartlett2016analysis}:  {\it However, these data, particularly the early stromatolite data [Pannella, 1972], should not be taken too seriously.
[Zahnle andWalker, 1987] Paleontologists Scrutton [1978] and Hofmann [1973] also found these data to
be unreliable and unsuitable for precise quantitative analysis.} These data have also been 
used in a crucial manner in the recent studies \citep{Mitchell2023a,wu2023day,baolength2023}.

Stromatolites are layered organo-sedimentary structures formed by the capture of sediments by microbial organisms, typically formed in shallow marine and lacurtrine environments. They are some of the oldest known forms of life, and are an important archive for studying Precambrian paleoenvironments. Studies of recent analogues have shown that daily rhythms of biological growth and binding of sediment can be preserved in stromatolites layering, as the microorganisms (algae) respond actively to daylight
\citep[e.g.][]{LoganRezak1964a,Monty1967a,Davies1970,Gebelein1969}. In addition, environmental fluctuations influence rates of growth and the supply of material, i.e., layering thickness, including tidal and seasonal variations
\citep{Gebelein1968,Pannella1976,pannella1972precambrianB}.
As such, investigations have been made of diurnal growth layers interpreted from fossil stromatolites that can be structured in larger tidal and seasonal banding;  these observations have in turn been used to estimate past LOD and the length of the lunar month. 

The analyses of Precambrian stromatolites by \citet{pannella1972paleontological,pannella1972precambrianB}  are a well-known example of such a study, and, as discussed are often used in long-term reconstructions of the past Earth-Moon system \citep[e.g.][]{bartlett2016analysis,baolength2023,Mitchell2023a,wu2023day}.
The three oldest, most critical LOD data that are often referenced are based on stromatolite sequences from the ca. 1.88-Ga Gunflint Formation \citep{fralick_age_2002}, the correlative Biwabik Formation (Lake Superior region), and the Paleoproterozoic Great Slave Supergroup (Northwestern Territories); the exact stratigraphic origin and thus age of the latter sample is unknown, but is usually placed around 2 Ga, following the compilation figure of \citet{williams2000geological}.
However, as emphasized in many studies \citep[e.g.][]{hofmann_stromatolites_1973,scrutton_periodic_1978,lambeck2005earth}, and by Pannella himself in the original publication, these stromatolite-based estimates are rarely, if ever, suitable for precise quantitative interpretation. The formation of daily laminae  depends on a fine environmental balance that can easily be disturbed, and therefore stromatolite  growth patterns are rarely considered to be complete due to periods of non-deposition  and/or post-depositional erosion by storms, for example. For this reason, counts or sequences of laminae should generally be interpreted as minimum estimates, and not as most likely values \citep{pannella1972paleontological,pannella1972precambrianB,lambeck2005earth}. Another source of uncertainty arises from ambiguities associated with determining the exact number of daily laminae per lunar monthly or annual bundle,  which is often characterized by a low reproductibility rate, and further challenged by a lack of independent temporal controls. This is a challenge that is, however, not unique to the stromatolite archive (tidal rhythmites, for instance, feature similar challenges). 
Concerning the stromatolite specimens studied by \citet{pannella1972paleontological,pannella1972precambrianB}, 
 we specifically note that his counts of the number of diurnal laminae per larger seasonal growth band yield a significantly lower interpreted number of days per year than the observed number of daily increments between lesser growth marks that are believed to indicate the synodic month, established from the same sequences. In addition,
\citet{Mohr1975a} arrived at a very different number and conflicting interpretation for time-equivalent specimens.
In conclusion, while stromatolites can provide useful insights in past tidal dynamics, they should typically not be considered as providing accurate or precise numerical values for past LOD reconstructions.

\subsection{Tidal rhythmites}
\label{sec:rhy}
Tidal rhythmites are laminae deposits related to semi-diurnal or diurnal tidal cycles that can occur in  estuaries or  deltas. Silty, and muddy sediment is carried by ebb tidal currents. These currents transport the sediment  in suspension through the main ebb channel to deeper offshore water, where it settles and forms graded layers. During slack water between tides, muddy caps can be deposited on the sandy laminae. The amount of sediment carried by ebb tidal currents and the effectiveness of tides in transporting and depositing sediment are directly related to the tidal amplitude \citep{williams2000geological}. In an ideal scenario, the analysis of the time series of the thickness of these laminae should allow to recover all tidal periodicities:
\begin{itemize}
\item
The lunar day, interval between two passage of the Moon at the meridian.  
\item
The lunar synodic month, which separates two full Moons, and is thus recovered by the recognition of spring  tides of large amplitude, occurring at syzygy, when the Moon and the Earth have same longitude, when seen from the Sun.  
\item
The tropical year, separating two passages of the Earth at the spring equinox (also called vernal equinox), possibly determined as the time of maximal tidal amplitude in the year. 
\item
Finally, if the record is sufficiently long, the nodal period of the Moon (18.6 yr at present) could be recognized \citep{walker1986lunar}.  
\end{itemize}

Although very promising, this method has its drawbacks. The locations where high-quality sequences of such laminae formations can be observed are rare. Moreover, they often led to divergent analyses when the deposits are analysed by different groups. The Weeli Wolli Formation in Western Australia, dated 2.45 Ga, was interpreted by \citep{walker1986lunar} on the basis of the lunar nodal cycle and led to an Earth-Moon distance of $51.9 \pm 3.3$ Earth radius ($R_E$), while the analysis of \citep{williams1989tidal,williams1990tidal} led to the much larger value of $54.6 \pm 1.8 \ R_E$, with the analysis of laminae couplets, grouped in synodic fortnightly increments, and annual cycles. 
Although the results of these two studies are still consistent within uncertainty (given the very large error bars), these estimates are based on two fundamentally different and mutually incompatible interpretations of the same layering patterns. 
In the same way, the analysis of the Elatina Formation in South Australia, dated 620 Ma, led \citet{williams1989tidal,williams1990tidal} to determine an Earth-Moon distance of  $58.16 \pm 0.30 R_E$ while \citet{sonett1998neoproterozoic}  derived 
$56.04 \pm 0.03 R_E$ from their analysis of the same sequence.  \citet{sonett1998neoproterozoic} also re-analyzed their previous determination of the Earth-Moon distance for the Big Cottonwood Formation in Utah, at 900 Ma, and found a value corresponding to an Earth-Moon distance of $55.06 \pm 1.44 R_E$, while their previous determination of  the same sequence was $57.1 R_E$ \citep{SonettKvale1996a}. We note that the new determinations from \citep{sonett1998neoproterozoic} are in  agreement with the new model of \citep{farhat2022resonant} (Fig.\ref{fig:f22}).
However, given the difficulties associated with interpreting Earth-Moon parameters from these tidal sequences, additional independent studies should be required to further verify the determination of \citep{sonett1998neoproterozoic}.

\subsection{Cyclostratigraphy}
\label{sec:cycl}
Due to the gravitational interactions of the Earth with  the other planets, the orbital plane of the Earth is moving  in a complex motion composed of a slow rotation (precession of the node) and a composition of nearly periodic motions that make the inclination of the Earth orbital plane oscillate \citep[e.g.][]{Laskar2020a}. This induces as well an oscillation of the tilt of the Earth (or obliquity, $\varepsilon$) of present amplitude  1.3 degrees around its averaged value. The variation of insolation on the surface of the Earth depends also on the precession of perihelion, and of the variation of eccentricity, which are dominated by the  so-called long eccentricity term of 405 kyr period and the short eccentricities of main periods 95 kyr and 124 kyr. Finally, the pull of the Moon and Sun on the equatorial bulge of the Earth induces a slow precessional motion of its spin axis at $50.475838$ arcsec/yr that corresponds to a  period of about 26kyr \citep{Laskar2020a}. Neglecting the eccentricity of the Earth and Lunar orbit, and the inclination of the Moon,  the lunisolar precession  can be expressed as $\alpha \cos \varepsilon$, where $\varepsilon$ is the obliquity and where the precession constant $\alpha$ is (eq.4.14 from \citep{Laskar2020a})

\def\moo{M} 
\def\ed{{E\!_d}}
\def\mo{m_\odot}
\def\o{\odot}
\def\temi{\Frac{3}{2}}

\begin{equation}
\alpha= \Frac{3}{2} G \,\left[ \Frac{\mo}{a_\o^3} + 
 \Frac{m_\moo}{\color{red}a_\moo^3}  \right]\Frac{\ed_0}{\gamma_0^2}{\color{red} \gamma}
 \label{eq:alpha}
\end{equation}
where G is the gravitational constant,  $\o$ refers to the Sun, and $M$ to the Moon, $m$ and $a$ are the masses and semi-major axes, $\gamma$ the Earth's spin angular velocity, $\gamma_0$ its present value, and $\ed_0$ the dynamical ellipticity at present. As $a_\o$ can be considered as constant, the precession constant $\alpha$  thus depends mostly on  the evolution of $a_\moo$ and $ \gamma$ which evolve in time under tidal dissipation (highlighted in red in equation (\ref{eq:alpha})).

The resulting changes in insolation drive climatic changes on Earth (astronomical climate forcing) that can be recorded in the Earth’s sedimentary archive. These sediments can today be studied \citep[e.g.][]{GradOgg2004a,Grad2012a,gradstein_2020,montenari_2018} and inform us on past astronomical changes. 

Over very long timescales, beyond 60 Ma,  the planetary orbital motions can no longer be predicted with accuracy \citep{LaskGast2011a,LaskFien2011b}, but for the Earth-Moon evolution, the tidal dissipation will dominate, and a reconstruction of the past evolution of the Earth-Moon distance can still be achieved. The variation of  $a_\moo$ and $\gamma$ will induce a change in the precession period that can be imprinted in the sedimentary record \citep{berger_stability_1992,meyers2018proterozoic}. In a reverse way,  the determination of the precession frequency  from the sedimentary record, and the use of a dynamical model that will link the semi-major axis $a_\moo$ to the angular spin velocity of the Earth $\gamma$ can allow to retrieve both $a_\moo$ and $\gamma$. This determination requires a time scale for the sedimentary record, which can be provided either by absolute radiometric age dating, but also, and more often by the use of the 405 kyr eccentricity period as a metronome for stratigraphic cycles (see \citep{Laskar2020a} and references therein). In recent years, this technique for determining the past state of the Earth-Moon system has made large progress, and  many groups have obtained converging results using various methods for the determination of the precession frequency \citep[e.g.][]{meyers2018proterozoic,lantink2022milankovitch,ZeedenLaskar2023a}. Moreover, these data are in good agreement with the tidal model of \citep{farhat2022resonant} (see fig.6 of \citep{farhat2022resonant} and fig.5 of \citep{ZeedenLaskar2023a}). 
Another crucial advantage of cyclostratigraphy, relative to, for example, stromatolites and tidal rhythmites, is the potential of independent age control by, for example, radioisotopic geochronology and integrated stratigraphic approaches. Using an integrated stratigraphic approach it is also possible to verify interpretations in time-equivalent sections that should have the same time-dependent astronomical signatures \citep[e.g.][]{OlseLask2019a,sinnesael_cyclostratigraphy_2019}. 
The coherence of these sets of data leads us to consider that they are the most robust among the geological proxies for the determination of the past precession frequency of the Earth and determination of Earth-Moon system parameters. 

\section{Discussion of the recently published results}

The solution of \citep{farhat2022resonant}  is in  agreement with the most recent 
determinations of tidal rhythmites \citep{sonett1998neoproterozoic} and with the recent cyclostratigraphic data \citep{meyers2018proterozoic,lantink2022milankovitch,ZhouWu2022a,ZeedenLaskar2023a} (Fig.\ref{fig:f22}). One could thus think that the fate of the thermal tides locking hypothesis was settled. But  the recent publication of the two papers \citep{Mitchell2023a,wu2023day} in major journals requires some additional discussion to clarify the situation.

\fige
\subsection{\citet{Mitchell2023a}}

In their compilation of  geological constraints of the Precambrian length of the day, 
 \citet{Mitchell2023a} have included most  of the available data. Their analysis is purely empirical. They search for the best linear fit, made by pieces over sequences of data.  One could wonder on the status of their fit, which is not continuous, as a piecewise linear model would be. The goal is thus not to find an empirical model  for the Earth-Moon evolution, but  to search for the best fitted trends  in  the LOD over extended periods. 
From these fits, they conclude to a probable lock of the LOD between 1 Ga and 2 Ga. 
When comparing to the results of  \citep{farhat2022resonant} (Fig.\ref{fig:mitchell}), one can observe that the stromatolites data at 1.88 Ga and 2 Ga from \citep{pannella1972precambrianB,pannella1972paleontological}  are essential for the conclusions of \citep{Mitchell2023a}. If these data, which are questionable as we discuss in section \ref{sec:stro}, are not taken into account, the fit will no longer lead to this locked value of the LOD between 1 and 2 Ga. 

It is also puzzling that the cyclostratigraphic point of \citep{grotzinger_cyclicity_1986} is nearly exactly on the  \citep{farhat2022resonant} curve (Fig.\ref{fig:mitchell}). 
It should  be noted, however, that the datum point of \citep{grotzinger_cyclicity_1986} was not originally given in that paper but was derived by \citet{Mitchell2023a}. \citet{grotzinger_cyclicity_1986} only proposed that there is  eustatic sea-level cyclicity within the Milankovitch frequency band recorded in platform carbonates from the Rocknest Formation, at a scale of $1-15$ m and possibly of $75-100$ m or  $75-200$ m. \citep{Mitchell2023a} then assumed that a  10 m cycle represents climatic precession and 87.5 m is related to short eccentricity. However, 
this interpretation is poorly constrained. 
Due to the large uncertainty of this data point, we  should consider this as a simple coincidence, until some new quantitative analysis in the spirit of \citep{meyers2018proterozoic} is performed on the same 1.89 Ga Rocknest Formation sample \citep{Mitchell2023a}. 

\figf
\subsection{\citet{wu2023day}}

In the recent work of \citet{wu2023day}, the authors presented a new analytical model of thermal tides to address the resonant locking hypothesis. The model's free parameters were constrained such that the resulting thermotidal torque drives an LOD history that best fits their compilation of LOD geological proxies. 
As previously, in figure \ref{fig:wu}, we  compare \citep{wu2023day} (black curve) with \citep{farhat2022resonant} (purple curve). Here again, one can see that the model of \citep{wu2023day} relies heavily on the stromatolite data of \citep{pannella1972paleontological,pannella1972precambrianB} to establish the resonance locking of the 
LOD\footnote{Note that these data points from Pannella \citep{pannella1972paleontological,pannella1972precambrianB} occur at a different age position in the figures $1-3$ of \citep{wu2023day}, compare to previous LOD compilations \citep[e.g.][]{williams2000geological,bartlett2016analysis,Mitchell2023a}. In particular, we note that the two closely spaced points at ca. 1.63 Ga most likely represent Pannella's analyses of the time-correlative Gunflint (1972a) and Biwabik (1972b) Formations dated at ca. 1.88 Ga \citep{fralick_age_2002}. However, used literature data were not provided in \citep{wu2023day} to verify this observation.}.
Moreover, \citep{wu2023day}  curve misses entirely the new cyclostratigraphic determinations of the Earth-Moon state at 2.46 Ga obtained by \citet{lantink2022milankovitch}  in the  Joffre Gorge, Australia.

\citet{wu2023day} elaborated a physical model to support their claims. Moreover,  the authors performed a suite of GCM (General Circulation Model) numerical simulations, using the LMD-G \citep{hourdin2006lmdz4} and PlaSim \citep{fraedrich2005planet} GCMs, to infer the Earth's paleo-temperature evolution that is required to generate the constrained history of the thermotidal torque. We dedicate the rest of this section to discuss the details behind the adopted model in \citep{wu2023day} and its predictions. 

\subsubsection{The modeled gravitational tides: artificial resonances?}
The dynamical evolution of the Earth's rotational motion in \citep{wu2023day} is driven by the luni-solar gravitational tidal torque and the solar thermotidal torque. For the former, the authors used the tidal history of \citep{webb1982tides}, where Laplace's Tidal Equations (the equations describing the tidal response of a shallow fluid layer; LTEs hereafter) were solved semi-analytically over a hemispherical equatorial ocean on the surface of the Earth. While the work of \citet{webb1982tides} was seminal in coupling LTEs with the dynamical evolution of the Earth-Moon system, 
the modeled history of the lunar orbit in \citep{webb1982tides} yielded a lunar formation epoch that is incompatible with the geologically constrained lunar age (see Fig.3 of \citep{webb1982tides}).  

To efficiently remedy the latter discrepancy, \citet{wu2023day} tweak the tidal dissipation history of \citet{webb1982tides} (see their Fig. S1) by normalizing it with a constant factor, such that the resultant orbital history of the Moon features its proper temporal origin. As a byproduct of this modeling choice, the authors have modified the spectrum of oceanic normal modes in such a way that tidal resonances are characterized with artificial amplitudes
\citep[see][]{green2017explicitly,daher2021long,farhat2022resonant}. 

Though the authors focus on modeling thermal tides,
gravitational tides remain the dominant driver of the Earth's rotational evolution, providing the background of the tidal torque upon which the thermotidal counterpart would significantly contribute only in the vicinity of the Lamb resonance. As such, since the authors are constraining the history of the total torque to fit a compilation of geological LOD proxies, an artificial modeled spectrum of gravitational tidal dissipation may yield an artificial spectrum of thermal tides. Namely, the resultant thermotidal history could be characterized by either an artificial timing of the Lamb resonance occurrence, an artificial amplitude of the Lamb resonance, or both.   
  
\subsubsection{Atmospheric thermal tides: model limitations}
\label{Sec:Wu_TT_analysis}
For the thermotidal contribution to the Earth's rotational history evolution, \citet{wu2023day} develop a simplified analytic model of thermal tides that is used to compute the thermotidal torque (Eqs. S28-S29 therein). The model is parameterized by a number of free parameters (16 in total) that are constrained such that the resulting thermotidal torque, added to the gravitational tidal torque, would drive an LOD history that fits the compilation of LOD geological proxies (see Figure \ref{fig:wu}).

The developed model essentially resembles a band-pass filter, similar to that developed in \cite{bartlett2016analysis}. It ignores the Coriolis force, which may be significant in the case of a fast rotator like the Earth, along with the vertical velocity of tidal waves. The model also assumes an isothermal structure of the atmosphere. This choice is classical in the literature of  atmospheric dynamics as it simplifies the mathematical framework of the rather complex theory \citep[e.g.,][]{chapman1969atmospheric,lindzen1972lamb,auclair2019generic}. However, for the Earth, atmospheric temperature measurements \citep[e.g., Figures 2.1-2.3 of][]{pierrehumbert2010principles} show that the massive troposphere (${\sim}80\%$ of atmospheric mass) controlling the tidal mass redistribution  is characterized by a negative temperature gradient. The latter is in fact closer to an idealised adiabatic profile than it is to an idealised isothermal profile. These modeling choices could deliver inaccuracies in the determination of the resonant period \citep{FarhatAuclair-Desrotour2023a}. However, this was somewhat compensated by the authors in modeling the resonant period as a free parameter that is constrained by the geological data.

The other essential quantity of interest is the amplitude of the thermotidal torque when the resonance is encountered. The latter is dependent on several variables, of which the least constrained in the case of the Earth is the rate of energy dissipation by the atmosphere. Namely, as the atmosphere is heated by the shortwave incident stellar flux and the infrared emission from the ground, it dissipates energy via multiple pathways including radiative cooling and frictional interactions with the surface. As it is difficult to properly model these mechanisms in the analytical theory, energy dissipation is usually modeled by a free parameter \citep[the parameter $Q_{\rm th}$ in the work of ][]{wu2023day}. This unconstrained parameter predominantly controls the amplitude and the spectral width of the resonant thermotidal torque and, consequently, the lifetime of the Lamb resonance and whether it was sufficient to counteract the gravitational tide.

Dissipative radiative transfer and atmospheric cooling, however, can be properly accommodated in GCM simulations. To that end, 
\cite{wu2023day} presents, to date, the first study that uses GCMs to simulate the Lamb resonance specifically for the Earth. Their results, using the two aforementioned GCMs,  estimate the dissipation parameter to be $Q_{\rm th}\approx10$, which would render the maximum amplitude of the torque insufficient for the LOD locking. For the LOD evolution, however, the authors used  values of $Q_{\rm th}$ that are one order of magnitude larger ($Q_{\rm th}\approx 100)$ such that the thermotidal torque would be sufficient to counteract the gravitational counterpart. Consequently, the used thermotidal torque is amplified by a factor of ${\sim}30$ relative to its present value\footnote{which means that the amplitude of the surface pressure anomaly is amplified by a factor of ${\sim}60$ as can be inferred from \cite{FarhatAuclair-Desrotour2023a}.}. The author's reasoning lies in the need for such a large thermotidal torque so that the LOD proxies, specifically the stromatolites in \citep{pannella1972paleontological,pannella1972precambrianB}, can be explained. This brings us back to Section \ref{sec:stro} in questioning the reliability of this data set as a robust constraint for informing dynamical models, especially when present with evidence from GCMs to the contrary. 
\figi

\subsubsection{The asymmetry of the Lamb resonance}

An interesting signature of the GCM simulations of the Lamb resonance in \citep{wu2023day} lies in the spectrum of the thermotidal torque shown in their Figure S4. We reproduce this spectrum in Figure \ref{fig:Wu_Spectrum}. The GCM spectrum, shown by the black dots, features an asymmetry in the peaks of the Lamb resonance whereby the two peaks of the torque around the resonance do not share the same amplitude. Namely, the accelerating part of the torque has an amplitude that is almost half that of the decelerating part. The former part is required to occur with a sufficient amplitude such that it counteracts the decelerating gravitational torque, but it appears from this spectrum to be reduced. The authors, however, ignored this signature present in the GCM simulations in favor of the spectrally symmetric Lamb resonance obtained from their analytical model, which is shown by the black curve in Figure \ref{fig:Wu_Spectrum}. 

In the recent work of \cite{FarhatAuclair-Desrotour2023a}, the authors propose that such an asymmetry can be obtained if one accounts for the thermal inertia budget in the ground and the lowermost atmospheric layer. Namely, due to the thermal inertia in these layers, the infrared heating of the atmosphere by the ground  becomes asynchronous with the incident solar flux. This delayed ground response is shown to be responsible for maneuvering the atmospheric tidal bulge in such a way that creates an amplitude asymmetry between the two peaks. In Figure \ref{fig:Wu_Spectrum}, we show by the red curve how the model of \cite{FarhatAuclair-Desrotour2023a} can properly explain the spectral asymmetry of the GCM-produced spectrum when taking into account the thermal inertia effects. It is important to note that the reduction of the positive peak of the torque goes hand in hand with the relative contribution of the ground in heating the atmosphere. Namely, the more abundant the greenhouse gases are in the atmosphere, which is predicted for the Precambrian from various geological proxies \citep[see e.g.,][]{catling2020archean}, the more the atmosphere would be prone to infrared thermotidal heating, and consequently the more the accelerating thermotidal torque would be reduced.

\subsubsection{The temperature problem} 

One naturally wonders how the discussed modeling limitations carry over to the model predictions. The resulting timing of the Lamb resonance occurrence requires a mean Earth temperature in the Proterozoic of $40-55^\circ$C, computed by the authors using the PlaSim GCM \citep[see Figure 7 of][]{wu2023day}. Though a warm climatic interval fits  evidence on a Proterozoic glacial gap \citep[e.g.,][]{hoffman2017snowball}, such extreme temperatures are in contrast with geochemical analysis using phosphates \citep[e.g.,][]{blake2010phosphate}, geological carbon cycle models \citep[e.g.,][]{sleep2001carbon,krissansen2018constraining}, numerical results of 3D GCMs \citep[e.g.,][]{charnay2020faint}, and the fact that solar luminosity was 10-25\% lower during the Precambrian \citep[e.g.,][]{gough1981solar}. Such extreme temperature estimates would also require elevated amounts of partial pressure of CO$_2$, reaching 200 mbar. This exceeds inferred estimates from various geochemical proxies \citep[see the review by][and references therein]{catling2020archean}. More importantly, however, this temperature increase will enhance the asynchronous thermotidal heating of the atmosphere by the ground in the infrared, as we describe in Section \ref{Sec:Wu_TT_analysis}. The latter would significantly attenuate  the peak of the tidal torque near resonance, rendering it insufficient for the LOD locking. The adopted model in \citep{wu2023day} did not account for this feedback effect. Moreover, atmospheric dissipation is also enhanced with the increased temperature, as discussed in \citep{FarhatAuclair-Desrotour2023a}, which has an additional effect of attenuating the resonant amplitude of the torque.

\subsection{\citet{baolength2023}}
\figg
While they do not invoke a thermal tides LOD trapping,  \cite{baolength2023} try also to reconciliate the LOD history with the \citep{pannella1972paleontological,pannella1972precambrianB}  data. This time, they propose that between 2 Ga and 1.5 Ga, a sudden growth of the Earth core led to a reduction of the spin rate of the Earth. In figure \ref{fig:bao}, we have compared their solution with \citep{farhat2022resonant} (purple curve). In this case again, the scenario depends crucially on the stromatolite data of \citep{pannella1972paleontological,pannella1972precambrianB}. If these data are removed, there is no longer the necessity to search for some peculiar scenario, and it can be recognized that the model of \citep{farhat2022resonant} fits most of the reliable geological data in a satisfactory manner. 

\subsection{\citet{FarhatAuclair-Desrotour2023a}}
\label{sec:far}
\figh
In their recent work, \citet{FarhatAuclair-Desrotour2023a} have revisited the atmospheric thermal tides computations for rocky planets and in particular for  the Earth. 
They have constructed an ab initio model of thermal tides on rocky planets with a neutrally stratified atmosphere. This feature is a major change with respect to previous models, where closed-form solutions are usually obtained assuming that the atmosphere is isothermal \citep{lindzen1967tidal,chapman1969atmospheric,lindzen1972lamb,auclair2019generic,wu2023day}.
Although both atmospheric structures provide appreciable mathematical simplifications, neutral stratification appears to better capture the negative temperature gradient that characterises the troposphere of the Earth, which contains most of the atmospheric mass. As the stability of stratification with respect to convection determines the strength of the Archimedean force exerted on fluid particles in the vertical direction, the neutral stratification approximation annihilates the buoyancy effects in the tidal response. The upward-travelling internal gravity waves are thus filtered out from the solution, leaving only the horizontal compressibility forces responsible for the propagation of the Lamb wave.
Another major change with respect to previous models 
\citep{lindzen1967tidal,chapman1969atmospheric,lindzen1972lamb,ingersoll1978venus,dobrovolskis1980atmospheric,auclair2019generic,wu2023day}
is the  consideration of heat absorption near the ground level and heat exchange between the atmosphere and ground that takes into account the thermal diffusive processes in the planetary surface layer. This model allows to obtain a closed-form solution for the frequency-dependent atmospheric tidal torque, and is in agreement with simulations using GCMs, both for Earth-like and Venus-like planets. Specifically, when applied to the Earth, their model predicts a resonant rotational period of 22.8 hr, which is in agreement with a recent analysis of pressure data on global scales 
\citep{sakazaki2020array}\footnote{This value is in  agreement with the $11.38\pm0.16\,{\rm hr}$  semi-diurnal period obtained by analyzing the spectrum of normal modes using pressure data on global scales \citep[see Table 1 of][first symmetric gravity mode of wavenumber $k=-2$]{sakazaki2020array}}, 
and the GCM prediction of \cite{wu_why_2023}. As such, the model predicts the occurrence of the Lamb resonance not in the Precambrian, but in the Phanerozoic, with an amplitude that is insufficient to counteract the luni-solar gravitational tidal torque. This does not exclude the occurrence of the crossing of the resonance, but as the luni-solar gravitational tidal torque  remains larger than the thermotidal torque, no LOD trapping  can  occur. The crossing of the resonance then results only in a small change of 
the Earth's rotational decceleration: the spin  decceleration rate is slightly increased before the resonance, and then reduced to roughly its previous value after the crossing of the resonance. 

The \citep{FarhatAuclair-Desrotour2023a} model depends on two parameters $(\sigma_0, \alpha_{\mathrm{A}})$ , which are the cooling frequency and opacity parameters, respectively. The frequency $\sigma_0$  is the inverse of the timescale associated with energy dissipation, which is assumed to result from radiative cooling in the model. The higher $\sigma_0$,  the more efficient is energy dissipation. The frequency $\sigma_0$  is thus tightly associated to the amplitude of the Lamb resonance (thus related to the parameter $Q_{\rm th}$ appearing in \citep{wu2023day}). The opacity parameter $\alpha_{\mathrm{A}}$ quantifies the fraction of incident Solar flux that is transferred to the atmosphere in the thermal tidal forcing. Consequently, this parameter takes its values between 0 (no tidal forcing) and 1 (maximal tidal forcing). Other model parameters are related to the present day atmospheric gas mixture and surface temperature, and they are therefore well constrained. For the  thermotidal torque to cancel the gravitational torque in the Precambrian  (and thus the LOD locking to occur in the Precambrian),  the $(\sigma_0, \alpha_{\mathrm{A}})$ pair needs to be below the associated black solid curve of (Fig.\ref{fig:fat}). Nevertheless, the observation of the present thermal atmospheric response and the constraint on the cooling frequency  $\sigma_0$ deduced from the cooling timescale estimated by \citet{leconte2015asynchronous}, impose to the $(\sigma_0, \alpha_{\mathrm{A}})$  pair to be inside the intersection of the two shaded regions, above the LOD lock threshold. Moreover, \citep{FarhatAuclair-Desrotour2023a} show that the crossing of the resonance, within the limitations of their analytical model, most probably occurred in the Mesozoic, and not in the Precambrian. In this case, the curve to consider is the dashed line, which leads to an even less probable LOD locking. 

\section{Conclusions}

The famous astronomer Carl Sagan (1934-1996)  used to say that extraordinary claims require extraordinary evidences, which is another version of the Occam's razor  in science, stating that simpler explanations should be preferred to more complicated ones, in absence of strong arguments. Adapted to the present problem of the evolution of the Earth-Moon distance and LOD, it can be expressed as: Do we need a LOD lock by thermal tides to explain the evolution of  the Earth-Moon system over its age?  Is there  a strong evidence for a LOD lock in the Precambrian? 

The answer to the first question is clearly negative, as the  \citep{farhat2022resonant} model provides a coherent scenario for the tidal history of the Earth-Moon system, without  the need of a resonant atmospheric tidal lock.

We have seen also how all papers advocating  for a LOD lock by thermal tides 
\citep{bartlett2016analysis,Mitchell2023a,wu2023day}, or the alternate scenario of a growing Earth's core \citep{baolength2023} rely critically on the stromatolite 
LOD estimates of \citet{pannella1972paleontological,pannella1972precambrianB}. 

However, as emphasied by Pannella himself, and by several authors that studied these data afterwards,  the validity of the stromatolite-based LOD estimates derived from the Paleoproterozoic Gunflint-Biwabik Formations and Great Slave Supergroup
should be questioned (see section \ref{sec:stro}). 
There is thus at present no reliable geological evidences to support these alternate scenarios. Moreover, \citep{Mitchell2023a} presented a cyclostratigraphy-based datum from cyclicities in the 1.9-Ga Rocknest Formation \citep{grotzinger_cyclicity_1986} which is not compatible with the stromatolite data of Pannella. In addition, the solution of \citep{wu2023day}  complies with the questionable  stromatolite data of \citet{pannella1972precambrianB,pannella1972paleontological} but not with the more reliable cyclostratigraphic data of 
\citep{lantink2022milankovitch}.

The crucial importance of the stromatolite data at 1.88 Ga and 2.0 Ga used in previous models is an important motivation for the search for alternate estimates of the LOD in this time interval or more generally in the interval of 1.5 Ga to 2.0 Ga. A preference should be given to high resolution cyclostratigraphic data, in the spirit of \citep{meyers2018proterozoic}. In particular, it would be very useful to re-analyze the cyclostratigraphic data at 1.9 Ga of  
\citep{grotzinger_cyclicity_1986}. More  generally, we would like here to emphasize the importance of taking dating (age) uncertainty into consideration when fitting variables through any type of empirical estimate derived from the geological records.

The two recent analytical, semi-analytical, and numerical  studies of \cite{wu2023day}  and \citet{FarhatAuclair-Desrotour2023a}, 
although providing opposite conclusions, have 
improved our understanding of the possibility of atmospheric thermotidal daylength locking in the Precambrian. 
The problem addressed in \cite{wu2023day} can be summarized as follows: two parameterized spectra of tidal torques, one gravitational and one thermal, are combined, and the parameters of the two counterparts are constrained such that the combined torque drives an LOD evolution that fits a compilation of geological proxies. Much can be appreciated in that work, especially in highlighting the significance of Earth-Moon angular momentum depletion via thermal tides, simulating the Lamb resonance for the Precambrian Earth using GCMs, and establishing, using GCMs, for the first time a correlation between the resonant period, temperature evolution, and atmospheric compositional variations. Moreover, in the limited sense, the adopted analytical models of \citep{wu2023day}, laying down the used spectra of the torques, appear to capture the fundamental dynamical behavior of oceanic and atmospheric tides. However, a closer look at the hierarchy of modeling assumptions in the two models and the constraints imposed by the geological proxies reveal that the story is much more nuanced. In short, stringent constraints on the LOD history were imposed by a subset of quantitatively questionable proxies, as we discuss in Section \ref{sec:stro}. The latter were combined with a spectrum of oceanic tides that does not physically describe the tidal response of the Earth's paleo-oceans. As such, the modeled atmosphere was constrained to encounter the Lamb resonance with an unrealistic amplitude for the torque and in a whistle-stop fashion such that the stromatolite records in the Proterozoic can be explained.


Using a neutrally stratified analytical model that is more adapted to the Earth's atmosphere than the usual isothermal model, and taking into account heat diffusion mechanism in the vicinity of the ground interface, \citet{FarhatAuclair-Desrotour2023a}  conclude that the amplitude of the Lamb resonance is not sufficient for the thermotidal torque to counteract the  luni-solar gravitational tidal torque in the Precambrian. Moreover, their analysis conclude that the crossing of the Lamb resonance most probably occurred in the Mesozoic, and not during the Precambrian (see section \ref{sec:far}), with even less chance of daylength locking.  Interestingly, the numerical GCM simulations of \cite{wu2023day} allow to strengthen the analytical model of \citet{FarhatAuclair-Desrotour2023a}  by probing the asymmetry of the Lamb resonance (Fig.\ref{fig:Wu_Spectrum}). These two studies should  provide the basis of future improved models for atmospheric thermal tides, but for now, we should conclude, both by the consideration of the geological evidences and by the comparison of theoretical models, 
that there is no clear arguments allowing to state that LOD locking occurred in the past history of the Earth.

\medskip\noindent
{\bf Figures credits.}
Figure 3 from \citep{Mitchell2023a}, figures 3 and S4 from \citep{wu2023day}, and figure 11  from 
\citep{baolength2023} were adapted according to CC BY licence (https://creativecommons.org/licenses/by/4.0/).

\medskip\noindent
{\bf Acknowledgements.}
This project has been supported by the French Agence Nationale de la Recherche (AstroMeso ANR-19-CE31-0002-01) and by the European Research Council (ERC) under the European Union’s Horizon 2020 research and innovation program (Advanced Grant AstroGeo-885250). MLL acknowledges funding from the Heising-Simons grant no. 2021 – 2797.
  
\phantomsection
\footnotesize
\setlength{\bibsep}{2pt plus 0.3ex}
\bibliographystyle{jxl}
\bibliography{TAT}
\end{document}